Zborn k.qxp  10.11.2016  7:40  Page 2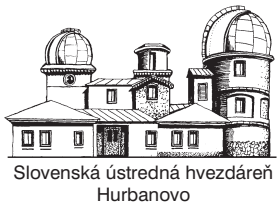

Slovenská ústredná hvezdáreň
Hurbanovo

Conference Dedicated to the 100th Anniversary of the Death of Dr. Nicolaus Thege-Konkoly,
and 145th Anniversary of the Founding of the Hurbanovo Observatory

Vydala Slovenská ústredná hvezdáreň Hurbanovo
Zostavovateľ: Mgr. Eduard Koči
Zodpovedný za vydavateľa: Mgr. Marián Vidovenec, generálny riaditeľ SÚH
Preklad do anglického jazyka: Silvia Kovács

ISBN: 978-80-85221-91-6

# Contents

# Heritage of Konkoly's Solar Observations: the Debrecen Photoheliograph Programme and the Debrecen Sunspot Databases

**Baranyi, T., Győri, L., Ludmány, A**.
*Heliophysical Observatory, Konkoly Observatory, MTA CsFK, Debrecen, Hungary*

**Abstract**
The primary task of the Debrecen Heliophysical Observatory (DHO) was to produce the detailed and reliable photographic documentation of the solar photospheric activity since 1958. This long-term effort resulted in various solar catalogues based on ground-based and space-borne observations. The DHO hosts solar-image databases containing heritages of two former Hungarian observatories. One of the sets of drawings was observed between 1872 and 1891 at the Ógyalla Observatory (now Hurbanovo, Slovakia) founded by Miklós Konkoly-Thege (1842–1916). We briefly summarize the history of the events that resulted in the longest photographic sunspot database available at the DHO at present, and we show the basic role of Dr. Miklós Konkoly--Thege in this achievement.

**History of Photoheliographic Databases**
The time-line of the main events was the following:
- 1843: The German amateur astronomer Heinrich Schwabe discovered the 11-year cycles of solar activity.
- 1845: The first clear image of the Sun was a daguerreotype taken by A. H. L. Fizeau and J. B. L. Foucault.
- 1852: Edward Sabine announced that the Schwabe's sunspot cycle was correlated very closely with the Earth's 11-year geomagnetic cycle. Astronomers became interested in observing the Sun.
- 1854: John Herschel argued the importance of obtaining daily photographic pictures of the Sun's disc, Royal Astronomical Society decided to support the building a photoheliograph for the Kew Observatory (also known as King's Observatory).
- 1857: Warren De la Rue produced the design for the Kew Photoheliograph, the first telescope specifically built to photograph the Sun.
- 1858: The systematic solar photographic observations started at the Kew Observatory, where Sabine controlled the geomagnetic and meteorological research.
- 1859: On 1 September, the magnetometers at the Kew Observatory recorded a brief but very noticeable jump in the Earth's magnetic field at exactly the same time as two amateur astronomers, R. C. Carrington and R. Hodgson, were the first to observe a flare on the Sun. It was the first observation of a space-weather event.
- 1860: The photoheliograph was briefly removed from the Kew Observatory to a site in Spain, where De la Rue used it to take the first good pictures of a total solar eclipse.
- 1861: The photoheliograph returned to the Kew Observatory, where the observers gathered 2778 white-light full-disc photographic observations for a full solar cycle between 1861 and

- 1872. The observatory gained renown and was well regarded for its three-fold activities (solar physics, geomagnetism, and meteorology).
- 1862: Konkoly finished his studies in Berlin, after that he visited a number of the large European observatories. It is probable that he knew the Kew Observatory as a good example for a successful observatory of three-fold activities with a strong emphasis on solar physics.
- 1871: Konkoly founded his observatory at Ógyalla *(http://www.suh.sk/history)*, and he equipped it with a heliograph in 1872. He started taking regular solar observations, designing and building solar telescopes and prominence spectroscopes. Figure 1 and 2 show two examples for his solar observations rich in details. Konkoly's meteorological station already worked at Ógyalla at this time, and later geomagnetic observations were also taken here. The Ógyalla Observatory gradually became a large astronomical, meteorological, and geomagnetic observatory similar to the Kew Observatory.

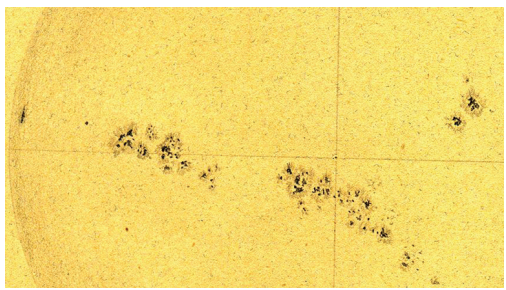

Figure 1. *A portion of the Konkoly's artistic full-disc solar drawing for Nov. 22, 1872.*

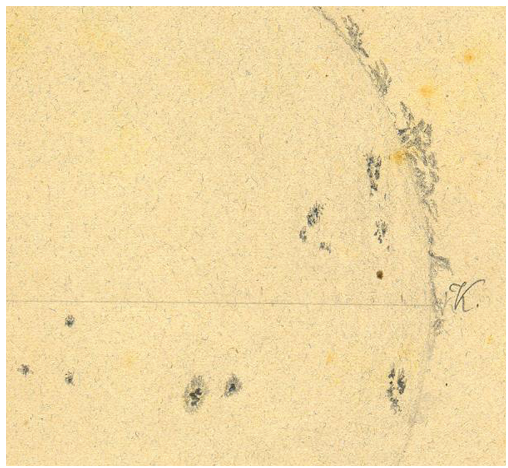

Figure 2. *A portion of the Konkoly's artistic full-disc solar drawing with prominences for Dec. 28, 1872.*

- 1872: George B. Airy, who was the Astronomer Royal at the Royal Observatory in Greenwich (RGO), regarded Kew as a rival over this decade. Airy achieved the transfer of Kew's photo-heliograph to Greenwich in August 1872. Solar observations taken here started to complete the observations of RGO's Magnetic Observatory.
- 1874: Start of the regular publication of the Greenwich Photoheliographic Results (GPR). GPR contained daily measurements of position and area data of sunspot groups (among other things). Konkoly knew the RGO well because he visited it a number of times.

**Konkoly's contribution to solar physics**
- 1874: Start of the regular publication describing the synoptic solar observations taken at Ógyalla by Konkoly or by employed observers (e.g. R. Kövesligethy, H Kobold).
- 1877: Konkoly installed a new telescope without a tube, specially designed for solar drawings.
- 1879: Start of the publication of celestial coordinates of sunspot groups including the observations for 1872–78.
- 1879: Konkoly helped in setting up the Haynald Observatory in Kalocsa, where regular solar observations started by Fenyi, Braun, and others. The observations taken here can be reckoned as indirect heritage of Konkoly. The set of solar drawings observed at the Haynald Observatory between 1880 and 1919 is also hosted at DHO.
- 1884: Installation of a new heliograph. Konkoly derived Wolf's relative sunspot number from all the observations available at Ógyalla.

- 1885–1917: Solar observations taken at Ógyalla's contribute to Wolf's sunspot relative number. Konkoly designed and built two photoheliographs, and he started photographing the solar photosphere in 1896. From 1907, the sunspots were counted to calculate Wolf's number in photographic images. The solar observations ceased in 1917 at Ógyalla, but the observations taken after 1891 are lost.
- 1898: Konkoly organised the 17. Ordentliche Versammlung der Astronomischen Gesellschaft meeting in Budapest (Figure 3 and 4). The event convinced the Hungarian government to take over the maintenance of the Ógyalla Observatory.

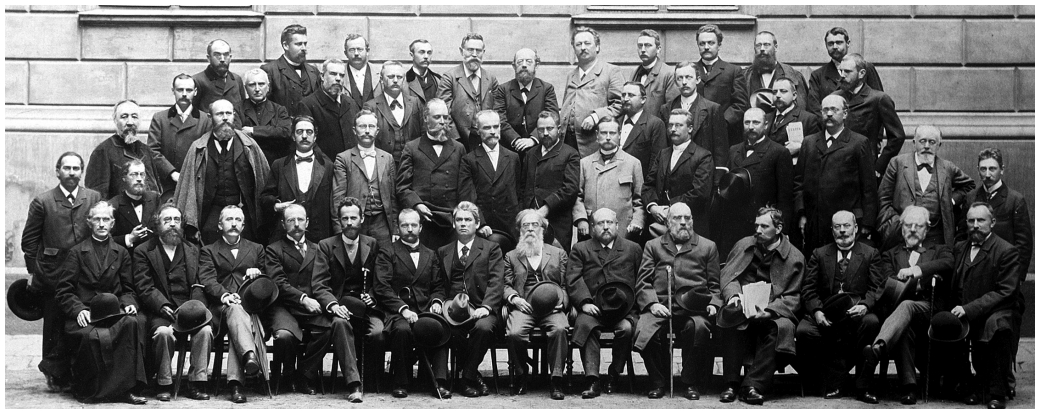

Figure 3. *Participants at the meeting 17. Ordentliche Versammlung der Astronomischen Gesellschaft, Budapest, 1898 (Image taken by György Klösz).*

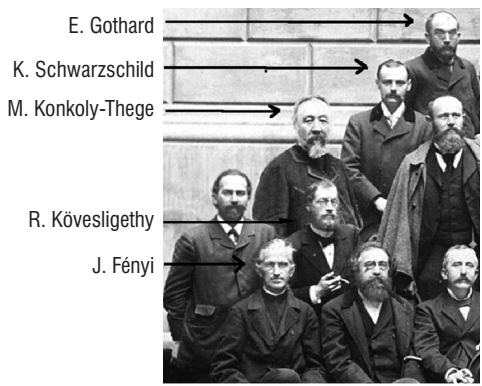

Figure 4. *Indication of some famous participants in an image enlarged from Figure 3.*

## Konkoly's instrumental heritage at the Debrecen Heliophysical Observatory

During the First World War, Konkoly's instruments were moved from Ógyalla to Budapest. In 1950, the solar observations were restarted by using the Konkoly's 5" photoheliograph at the Konkoly Observatory in Budapest. In 1958, the solar department of the Konkoly Observatory moved to Debrecen, became independent from the Konkoly Observatory, and it formed the Debrecen Heliophysical Observatory (DHO) of the Hungarian

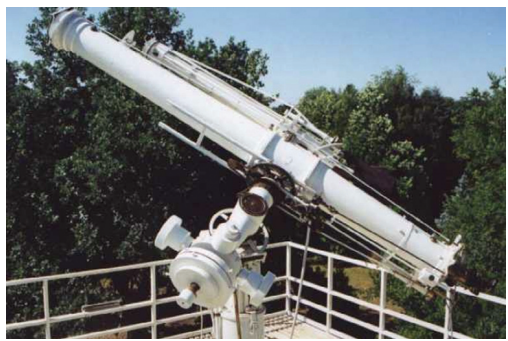

Figure 5. *Konkoly's instrumental heritage at the Debrecen Heliophysical Observatory. There were two of Konkoly's telescopes with Merz-objectives. The 5" photoheliograph was the tool of gathering the full-disc white-light observations while the 10" telescope was used for visual inspection of the seeing.*

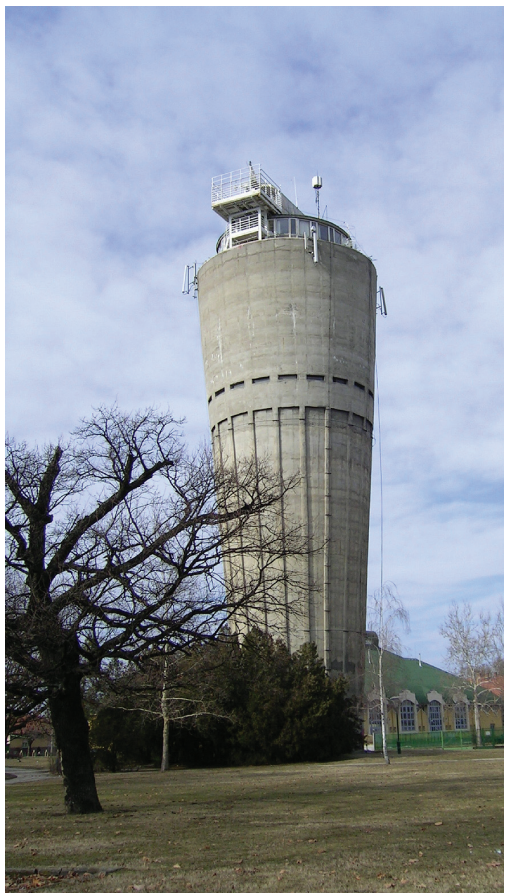

Figure 6. *The water-tower at Gyula with the 6" telescope of Konkoly at the Gyula Observing Station on its top.*

Academy of Sciences. Here, three of Konkoly's instruments continued serving the solar physics research. In Debrecen, there were two photoheliograps with Merz-objectives. Konkoly's 5" photoheliograph was used for synoptic full disc white light observations, and his 10" telescope was used for visual inspection of the seeing until 2008 (Figure 5) when the repair of its moving parts started. Between 1972 and 2006, a 6" photoheliograph was used for synoptic full-disc white-light photographic observations at the Gyula Observing Station of the DHO (Figure 6). In 2006–2007, the telescope at Gyula was renewed and equipped with a CCD camera started working in 2008. This camera worked until March 2011 when it was replaced with a new, larger camera by the beginning of 2012. In May 2013, the rebuilding and automation of the Gyula telescope started. Unfortunately, all attempts for the renewal of the telescopes failed because of the lack of the sufficient resources. In 2016, the observing sites at Gyula and Debrecen were closed, and the telescopes were moved to the Konkoly Observatory in Budapest to be installed in a future astronomical exhibition.

The instruments at Debrecen and Gyula gathered the largest full-disc white-light observational material between 1958–2013. The types of photographic plates were Kodak (Kodalith) or Agfa Gevaert, their size was 14×14 cm. The filter wavelength range was λ 5500 Å with Δλ = 50 Å. The diameter of solar disc was ~ 11 cm (Debrecen) and ~ 10 cm (Gyula) in the photographic observations. The size of the Gyula CCD was 4000×4000 pixels in the period 2008–11, and it was 8000×6000 in 2012–13. The number of photoheliograms taken at the two observing site is ~150,000–200,000. Figure 7 and 8 show an observation and its enlarged portion as a sample.

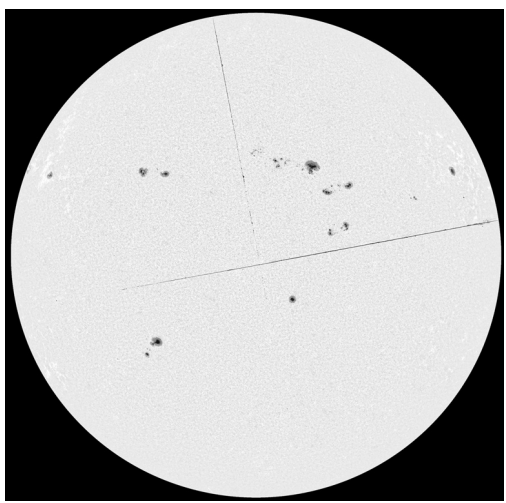

Figure 7. *One of the photoheliograms taken at Gyula on 16 June 2000.*

Figure 8. *An enlarged portion of the solar disc in Figure 7 showing one of the active regions. The details of the structure of the sunspots and the small pores that can be seen show the high quality of the observation.*

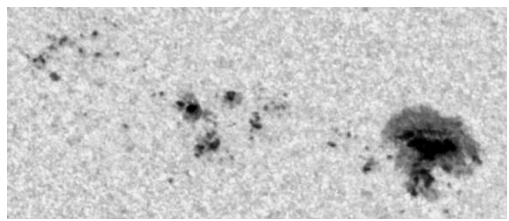

The specific features of the observational material are the followings:
– Series of 3 observations for calibration and averaging were taken within 10–15 minutes.
– Several series were taken in each day depending on weather (the number of series depended also on solar activity level and season (at least 2 series in winter and 3 series in summer were regularly taken a day).
– Visual checking of seeing for the best achievable quality was applied.
– Special attention was paid to the precision of position data, in average, it was ~0.1 heliographic degrees (in other datasets it is usually about one degree).

The photoheliographs of Konkoly were dedicated to a long-term solar observational programme at Debrecen and Gyula (Hungary) until 2016 when the observational facilities were closed. The programme started about four decades after Konkoly's death and during more than half a century it produced the most detailed photographic documentation of the solar photospheric activity.

**Scientific connection between Ógyalla Observatory and RGO via Debrecen Observatory**

In 1976 the termination of GPR was announced at the IAU General Assembly (GA) in Grenoble then the DHO undertook this task. The work to publish the Debrecen Photoheliographic Results started at DHO. In 1982, the DHO became a department of the Konkoly Observatory (KO) again.

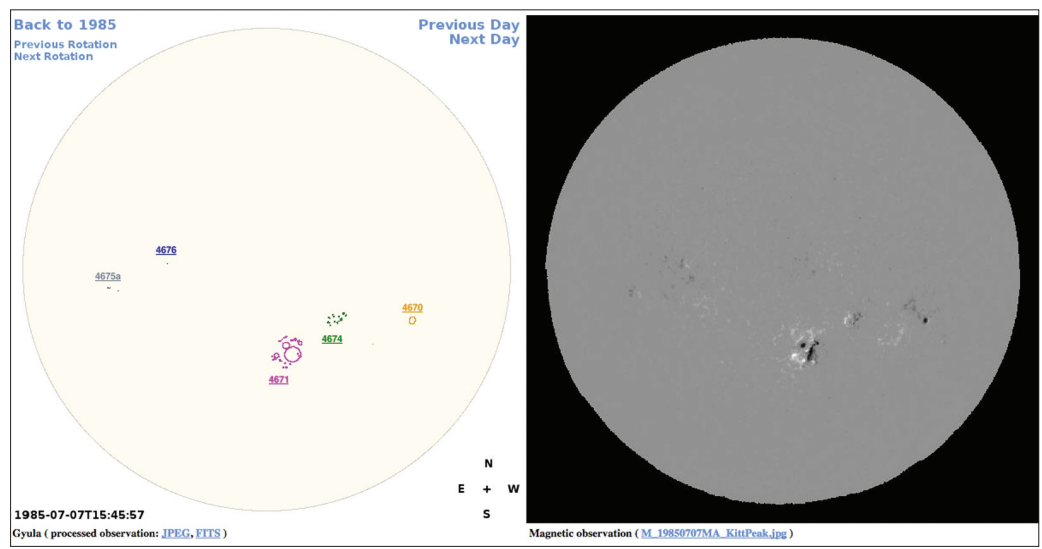

Figure 9. *Example of a web page showing the available information in the DPD for a given date available at http://fenyi.solarobs.unideb.hu/DPD/. Left panel: Schematic drawing of sunspots and sunspot groups visualizing the data for July 7, 1985. It is created from the position and area data of spots, derived from the observation indicated at the bottom of the panel. The spots (and pores) are represented by ellipses visualizing approximately the spot roughly as a projection of a circle on a sphere onto a plane. The centroid of the ellipse is at the position of the visible centroid of the spot. The area of the ellipse is the projected whole spot area. Right panel: Magnetogram for that day.*

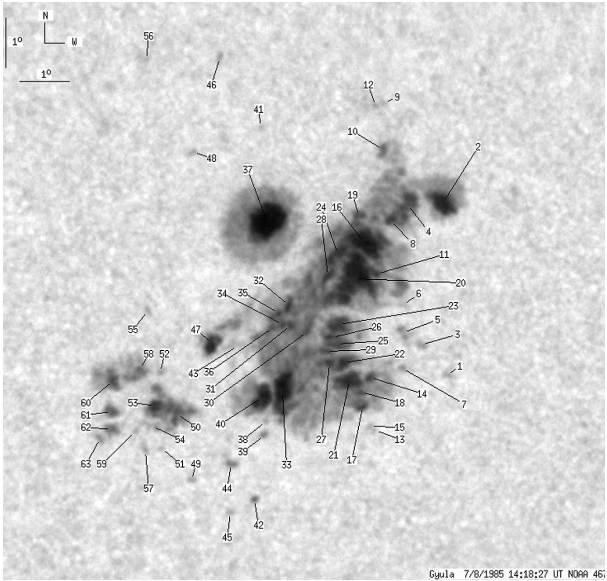

Figure 10. *Example for an image of active region on the website of DHO. The image can be seen after opening the link at AR NOAA 4671 in Figure 9. The spatial scale of the image of sunspot group is shown by line segments of one heliographic degree long in the upper left corner. All of the numerical data of the numbered features of the sunspot group can be seen in the table below the image on the site opened.*

Figure 11. *Example for a web page of GPR data appended with Konkoly's observation for Oct. 2, 1874 available at http://fenyi.solarobs.unideb.hu/GPR/. In the left panel, the graphical presentation of the GPR data is similar to that of the DPD data in Figure 9. In the middle panel, the full-disc drawing of solar photosphere by Konkoly can be seen. The Ógyalla drawings are East-West side-reversed, thus two options are available for displaying them. The original view is suitable for reading the text in the image (date or group number). The side-corrected view is suitable for comparing with the GPR data. If someone moves the mouse over the file names below the image, the versions of the image can be swapped. If one clicks on a file name, the image opens in a pop-up window in which the image can be enlarged as it can be seen in the right panel. If someone wants to compare the observation with the schematic drawing of GPR data, it has to be taken into account that the orientation of the original observation differs from that of the schematic drawing. In the Ógyalla images, the terrestrial North is at the top while in the schematic GPR drawings the solar North is at the top. The small orientation figure in the middle of the page between the original and schematic drawings helps in comparing the orientation of the two images showing the position angle $P_0$ of the solar North (N in red) measured eastwards from the terrestrial North point (N in black) of the solar disc.*

The DHO was encouraged to speed up the publication of sunspot data, therefore a separate project was launched to produce the Debrecen Photoheliographic Data (DPD) in 1993. The main contributors to DPD were Gyula and Debrecen but observations taken by several contributing observatories were used to ensure the full coverage of the years with daily data. The DPD became the most detailed long-term sunspot catalogue of the post GPR era. By the end of 2014, the DPD team (permanent members were T. Baranyi, L. Győri, and A. Ludmány) produced 41 volumes of DPD for 1974–2014 published during 23 years. Finally, the revised version of GPR has been converted to DPD format and they were published online in a unified form. In this way, the longest homogeneous photographic sunspot database was created with daily cadence starting at 1874. Full-disc white-light images and magnetic observations are appended to provide the morphological and polarity information available concerning the sunspots. All of the data and images are accessible by ftp to provide an easy bulk download, but a user-friendly HTML-platform is also provided for the the entire dataset at *http://fenyi.solarobs.unideb.h*u (see examples in Figures 9, 10, and 11). The contribution of the Konkoly's heritage to these databases is two-fold. On one hand, the Konkoly's telescopes provided photographic observational input into the DPD taken at Debrecen and Gyula during four decades. On the other hand, the historical solar disc drawings were appended to the revised GPR data. Documentation of the sunspot activity by using the historical solar drawings taken at Ógyalla for 1872–1891 (2140 observations by Konkoly et al.) and those of taken at Kalocsa for 1880–1919 (5904 observations by Fényi et al.) (see also at *http://fenyi.solarobs.unideb.hu/HHSD.htm*) provides a revitalization of Konkoly's heritage. These drawings, which are rich in details, may help us to reveal the structure of sunspots and sunspot groups in those years when there are no photographic images available. The presentation of the observations on-line ensures world-wide access to the observations taken at Ógyalla in agreement with the intentions of Konkoly. The observations ceased at DHO but the heritage of Konkoly is more living than ever before.